\documentclass[preprint,prd,11pt,nofootinbib]{revtex4-1}

\pdfoutput=1

\usepackage{graphicx}
\usepackage{graphics}
\usepackage[mathscr]{eucal}
\usepackage{amssymb}
\usepackage{amsmath}
\usepackage{xspace}
\usepackage{listings}
\usepackage{ulem}
\usepackage{color}
\usepackage{feynmp}
\usepackage{hyperref}

\definecolor{darkgreen}{rgb}{0,0.5,0}

\newcommand\SARAH{{\tt SARAH}\xspace}

\newcommand\SPheno{{\tt SPheno}\xspace}

\newcommand\NMSPEC{{\tt NMSPEC}\xspace}
\def\mueff{\mu_{\rm{eff}}}
\newcommand{\Veff}{V_{\text{eff}}}

\newcommand{\AddrBonn}{%
Bethe Center for Theoretical Physics \& Physikalisches Institut der
 Universit\"at Bonn, \\
53115 Bonn, Germany }

\newcommand{\AddrParis}{%
1-- Sorbonne Universit\'es, UPMC Univ Paris 06, UMR 7589, LPTHE, F-75005, Paris, France \\
2-- CNRS, UMR 7589, LPTHE, F-75005, Paris, France}

\newcommand{\AddrCERN}{%
Theory Division, CERN, 1211 Geneva 23, Switzerland
}

\begin{document}
\title{On the two-loop corrections to the Higgs masses in the NMSSM}
\author{Mark D. Goodsell}
\email{goodsell@lpthe.jussieu.fr}
\affiliation{\AddrParis}

\author{Kilian\ Nickel}
\email{nickel@th.physik.uni-bonn.de}
\affiliation{\AddrBonn}

\author{F.\ Staub}
\email{florian.staub@cern.ch}
\affiliation{\AddrCERN}

\keywords{supersymmetry, Higgs masses, NMSSM}


\preprint{Bonn-TH-2014-16, CERN-TH-2014-231}
\begin{abstract}
We discuss the impact of the two-loop corrections to the Higgs mass in the NMSSM
beyond $O(\alpha_S(\alpha_b + \alpha_t))$. For this purpose we use the combination 
of the public tools \SARAH and \SPheno to include all contributions stemming from superpotential parameters. 
We show that the corrections in the case of a heavy singlet
are often MSSM-like and reduce the predicted mass of the SM-like state by about 1~GeV 
as long as $\lambda$ is moderately large. For larger values of $\lambda$ the additional corrections
can increase the SM-like Higgs mass. 
If a light singlet is present the additional corrections become more important even for smaller values 
of $\lambda$ and 
can even dominate the ones involving the strong interaction.  In 
this context we point out that important effects are not reproduced quantitatively when only including 
$O((\alpha_b+\alpha_t+\alpha_\tau)^2)$ corrections known from the MSSM.
\end{abstract}
\maketitle


\nopagebreak

\section{Introduction}
The discovery of the Higgs boson at the Large Hadron Collider (LHC) 
has completed the standard model (SM) of particle physics \cite{Chatrchyan:2012ufa,Aad:2012tfa}.
However, there are some hints that the SM cannot be the final theory for particle physics: 
it does not provide a dark matter candidate, cannot explain the observed baryon asymmetry or neutrino masses, 
and suffers from the hierarchy problem. There are different options to solve all of these problems.
Supersymmetry (SUSY) is still the most popular among them, see Ref.~\cite{Nilles:1983ge,Martin:1997ns} and references therein. 
In the past the focus has mostly 
been on the minimal supersymmetric standard model (MSSM) but recently the interest 
in non-minimal SUSY models has grown, mainly for two reasons: (i) no hints for SUSY particles have so far been detected at the LHC;
(ii) the observed Higgs mass of $m_h \simeq 125$~GeV is rather close to the upper limit of 132~GeV 
possible in the MSSM for moderate SUSY masses including three-loop corrections \cite{Feng:2013tvd,Kant:2010tf}. 
Together these lead to the question of how natural the MSSM is. The situation becomes significantly improved if 
a singlet superfield is added to the particle content of the MSSM: the interactions with the singlet can 
give a push to the Higgs mass at tree-level \cite{Ellwanger:2009dp,Ellwanger:2006rm}
which reduces the necessary fine-tuning 
significantly \cite{BasteroGil:2000bw,Dermisek:2005gg,Ellwanger:2011mu,Ross:2011xv,Ross:2012nr,Ross:2012nr,Kaminska:2014wia}. 
The NMSSM has a rich collider phenomenology \cite{Dreiner:2012ec} and it has been shown recently that light singlinos could help to hide SUSY
at the LHC \cite{Ellwanger:2014hia}. The most studied singlet extension is the next-to-minimal supersymmetric standard model 
\cite{Ellwanger:2009dp} which introduces a $\mathbb{Z}_3$ to forbid all dimensionful parameters in the superpotential. 

To confront the NMSSM with the measured Higgs mass a precise calculation is necessary. Therefore, 
much effort has been expended on a full one-loop calculation in the $\overline{\mathrm{DR}}'$-scheme \cite{Degrassi:2009yq,Staub:2010ty} as well as different on-shell 
schemes \cite{Ender:2011qh,Graf:2012hh}. 
At the two-loop level the dominant $\alpha_S(\alpha_b+\alpha_t)$ have been available for a few years \cite{Degrassi:2009yq},
but two-loop corrections involving only superpotential couplings such as Yukawa and singlet interactions have not been 
available so far. 
Up until now the state-of-the art codes to calculate the Higgs mass written specifically for the NMSSM are \NMSPEC \cite{Ellwanger:2006rn},
{\tt Next-to-Minimal SOFTSUSY} \cite{Allanach:2001kg,Allanach:2013kza,Allanach:2014nba}, and {\tt NMSSMCALC}  \cite{Baglio:2013iia}.
The first two codes use the known \emph{MSSM} $(\alpha_b+\alpha_t+\alpha_\tau)^2$ and $\alpha_s(\alpha_t+\alpha_b)$ corrections, while {\tt NMSSMCALC} includes only $(\alpha_s \alpha_t)$ at two-loop. 
However, in Ref.~\cite{Goodsell:2014bna} an extension of the Mathematica package \SARAH 
\cite{Staub:2008uz,Staub:2009bi,Staub:2010jh,Staub:2012pb,Staub:2013tta} has been presented 
which can write a two-loop calculation for \SPheno \cite{Porod:2003um,Porod:2011nf} of CP-even Higgs masses in models beyond the MSSM with a similar precision as known for the MSSM \cite{Brignole:2001jy,Degrassi:2001yf,Brignole:2002bz,Dedes:2002dy,Dedes:2003km}. 
To be more precise, a lot of effort has been (and will be) made to take the Higgs mass precision in the MSSM beyond the effective potential \cite{Martin:2004kr,Martin:2007pg,Borowka:2014wla,Degrassi:2014pfa}, including RGE improvement and diagrammatic calculations. Advanced techniques are implemented in {\tt FeynHiggs} \cite{Hahn:2013ria,Heinemeyer:2007aq,Hahn:2005cu,Heinemeyer:1998yj}.

This calculation is based on the effective potential approach and uses generic results 
presented in Ref.~\cite{Martin:2001vx}. In this work we discuss the importance of these `full' two-loop calculations in the NMSSM, pointing out when it is not sufficient to use the known MSSM corrections. As explained later the calculation performed by \SPheno/\SARAH does not include corrections stemming from $g_1$ and $g_2$ but we will denote them nevertheless by `full' for brevity since all corrections stemming from superpotential parameters are included. This corresponds to the precision of state of the art calculations widely used in the MSSM, with the possible exception of corrections to the mass of the $Z$ boson which feed into the determination of the electroweak scale.

We proceed as follows: in sec.~\ref{sec:method} we fix our conventions and present details of the two-loop 
mass calculation. In sec.~\ref{sec:numerics} we show the numerical impact of the full two-loop calculation before we conclude in sec.~\ref{sec:conclusion}. The reader may refer to the appendix for some details of the calculation of the tree and one-loop Higgs mass matrices in our conventions; in particular, we describe there the treatment of the would-be Goldstone bosons, which provide some technical challenges for the calculation of the Higgs mass in the effective potential approach.

\section{Calculation of the Two-loop Higgs masses in the NMSSM}
\label{sec:method}
In this section we fix our notation and discuss the procedure for the 
calculations of the Higgs masses at two loops. Note that we shall focus throughout on the `NMSSM' as the theory invariant under a $\mathbb{Z}_3$-symmetry, referring only briefly to more general variants. 

\subsection{Superpotential and soft SUSY breaking terms of the NMSSM}
In the NMSSM, the particle content of the MSSM is extended by a gauge singlet superfield $\hat{S}$. 
We are using in the following the convention that all superfields are denoted by a 'hat', the SUSY partners
carry a 'tilde' and the SM fields come without any decoration. 
All interactions in the superpotential consistent with an underlying $\mathbb{Z}_3$ symmetry are given by
\begin{equation}
 W_{\rm{NMSSM}} = -\hat{H}_u  \hat{Q} Y_u \hat{U}^c + \hat{H}_d\hat{Q} Y_d  \hat{D}^c 
                + \hat{H}_d\hat{L} Y_e  \hat{E}^c  + \lambda \hat{H}_u \hat{H}_d \hat{S} 
                + \frac{1}{3} \kappa \hat{S} \hat{S} \hat{S} .
\label{eq:superpotential}
\end{equation}

The corresponding soft-terms read
\begin{eqnarray}
\nonumber
 \mathscr{V}_{\rm SB,2} &=&  m_{H_u}^2 |H_u|^2 + m_{H_d}^2 |H_d|^2+ m_S^2 |S|^2 
    + \tilde{Q}^\dagger m_{\tilde{Q}}^2 \tilde{Q}  +\tilde{L}^\dagger m_{\tilde{L}}^2 \tilde{L}
          + \tilde{D}^\dagger m_{\tilde{D}}^2 \tilde{D}
          + \tilde{U}^\dagger m_{\tilde{U}}^2 \tilde{U} + \nonumber \\
 && \hspace{1cm}  + \frac{1}{2}\left(M_1 \, \tilde{B} \tilde{B} + M_2 \, \tilde{W}_a \tilde{W}^a  + M_3 \, \tilde{g}_\alpha \tilde{g}^\alpha + h.c.\right) \\
\mathscr{V}_{\rm SB,3} &=& - H_u \tilde{Q}  T_u\tilde{U}^\dagger
                      +H_d  \tilde{Q} T_d \tilde{D}^\dagger 
                      + H_d \tilde{L} T_e \tilde{E}^\dagger 
                      + T_\lambda H_u H_d S  + \frac{1}{3} T_\kappa S S S
\end{eqnarray} 
After electroweak symmetry breaking (EWSB), the singlet is split into its CP even and odd component as well as 
a vacuum expectation value (VEV), just like the neutral parts of the Higgs doublets:
\begin{eqnarray}
\label{VEVsinglet}
 S &=& \frac{1}{\sqrt{2}}  \left(\phi_s + i \sigma_s + v_s \right) \, ,\\
 H_i^0 &=& \frac{1}{\sqrt{2}}  \left(\phi_i + i \sigma_i + v_i \right) \hspace{1cm} i=d,u \,.
\end{eqnarray}
We choose a phase convention such that all VEVs are real. The VEV of the singlet triggers effective $\mu$- and $B\mu$-terms
\begin{equation}
\mueff = \frac{1}{\sqrt{2}} \lambda v_s,\qquad B_{\mu,\rm{ eff}} = \frac{1}{\sqrt{2}}T_\lambda v_s + \frac{1}{2} \kappa \lambda v_s^2.
\end{equation}
We shall treat $\mu_{\rm{eff}}$ as an input parameter from which we extract $v_s$. The tree-level mass matrices for the neutral scalars and pseudoscalars are given in our notation in the appendix (eqs.~(\ref{eq:treescalarmatrix},\ref{eq:sigmamasses}), and the tadpole equations are given in equations (\ref{eq:tadpoleD} -- \ref{eq:tadpoleS}). Throughout we shall make the choice of solving the three equations for $m^2_{H_d}, m^2_{H_u}$ and $m^2_{S}$, leaving the input parameters to be 
$$(  \lambda,\mueff, T_\lambda, T_\kappa)$$
in addition to the parameters of the MSSM. It will sometimes be convenient in the following to define
$$ T_\lambda \equiv \lambda A_\lambda, \qquad T_\kappa \equiv \kappa A_\kappa.$$

We shall distinguish between two regimes of principal interest for the model: ones with a ``heavy'' singlet and ones with a ``light'' singlet, by which we mean respectively a singlet heavier than the standard-model-like Higgs, and  of comparable or smaller mass.

\subsection{Two-loop self-energies}
\label{sec:twoloopcalc}
The calculation of the two-loop self-energies is performed with the public 
tools \SARAH and \SPheno. The method applied by these tools is the one presented in 
Ref.~\cite{Martin:2002wn}: the effective potential is calculated at the two-loop level using the 
generic expressions of Ref.~\cite{Martin:2001vx}. The tadpole contributions and 
self-energies are calculated by taking the 
first and second derivative of the two-loop effective potential $\Veff^{(2)}$
\begin{eqnarray}
\delta t^{(2)}_i &=& \frac{\partial \Veff^{(2)}}{\partial v_i}  \\
\Pi_{h_i h_j}^{(2)}(0) &=& \frac{\partial^2 \Veff^{(2)}}{\partial v_i \partial v_j} 
\end{eqnarray}
with $i=d,u,s$. We give details of the tree-level mass matrices and further details of the procedure for calculating the loop-level masses in the appendix.

In \SARAH/\SPheno there are currently two possibilities to take the derivatives: one can numerically calculate the derivative of the entire
potential; or one can take the derivative of the potential with respect to the masses analytically and numerically evaluate the derivatives  
of the masses and couplings with respect to the VEVs. Since the contributions to the two-loop Higgs mass for the NMSSM presented below are new there are no extant codes or calculations with which to compare the full results, and so we have compared the results for the two different methods to ensure that there was agreement. In addition, there is a third possible method of calculation based on a fully analytical diagrammatic approach to the Higgs mass calculation which will be presented elsewhere \cite{POLEInProgress}, which has been implemented in \SARAH/\SPheno and will be made available in a future release. Since this is a rather independent calculation that does not rely on computing the effective potential itself the comparison with the other two methods is a non-trivial check; we have performed this and found excellent agreement.

\begin{figure}[hbt]
\includegraphics{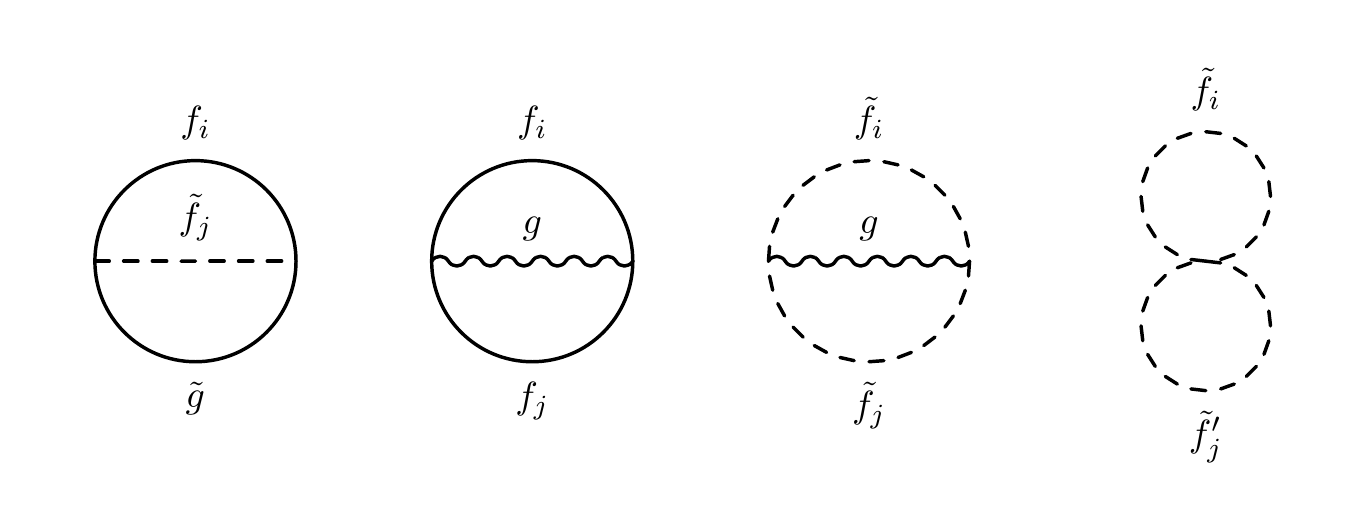}
\caption{Feynman diagrams contributing to  $\alpha_S(\alpha_b + \alpha_t)$ corrections in 
the effective potential at the two-loop level. The corrections involve SM fermions ($f_i,f'_i \equiv d_i,u_i,l_i,\nu_i$) 
and SUSY sfermions ($\tilde{f}_i,\tilde{f}_i'\equiv\tilde{d}_i,\tilde{u}_i,\tilde{l}_i,\tilde{\nu}_i$).
A sum over all possible flavour combinations 
is assumed, but all CP and charge violating diagrams are not considered. }
\label{fig:DiagramsAlphaS}
\end{figure}

\begin{figure}[hbt]
\includegraphics{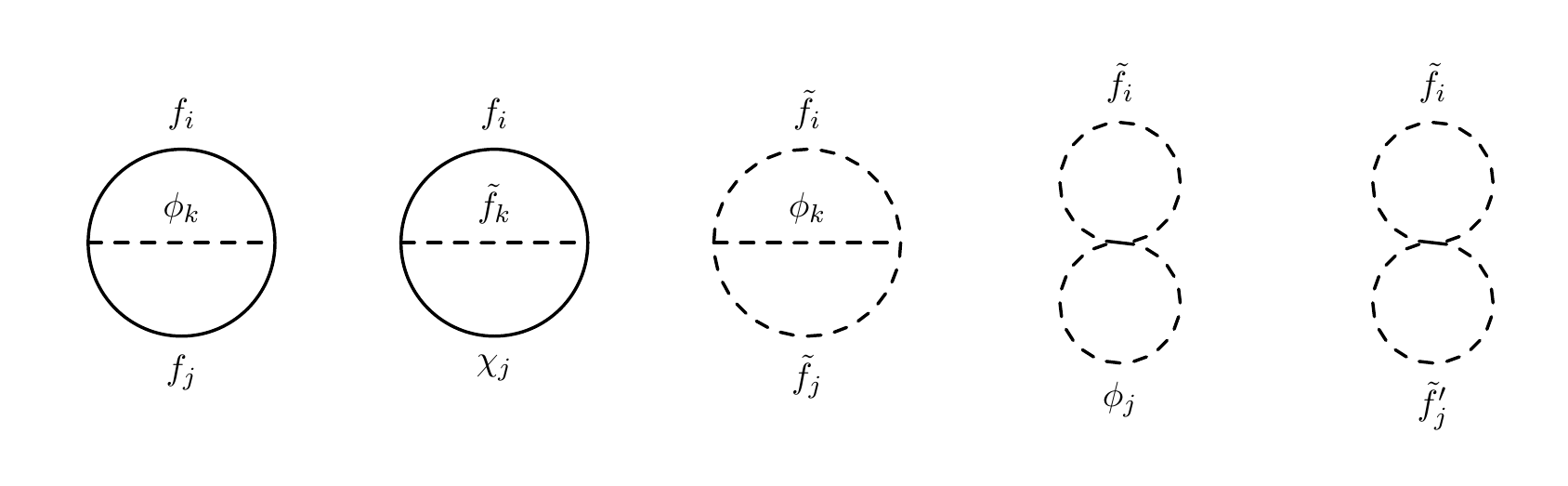}
\caption{Feynman diagrams involving SM Yukawa couplings which contribute to the effective potential at the two-loop level. 
The corrections involve SM fermions ($f_i,f'_i \equiv d_i,u_i,l_i,\nu_i$), 
SUSY sfermions ($\tilde{f}_i,\tilde{f}_i'\equiv\tilde{d}_i,\tilde{u}_i,\tilde{l}_i,\tilde{\nu}_i$), 
neutralinos/charginos  $\chi_i \equiv \tilde{\chi}^0_i, \tilde{\chi}^+_i$,
and Higgs particles $\phi_i \equiv h_i,A_i^0,H_i^+$. 
A sum over all possible flavour combinations 
is assumed, but all CP and charge violating diagrams are not considered. }

\label{fig:DiagramsYukawa}
\end{figure}

\begin{figure}[hbt]
\includegraphics{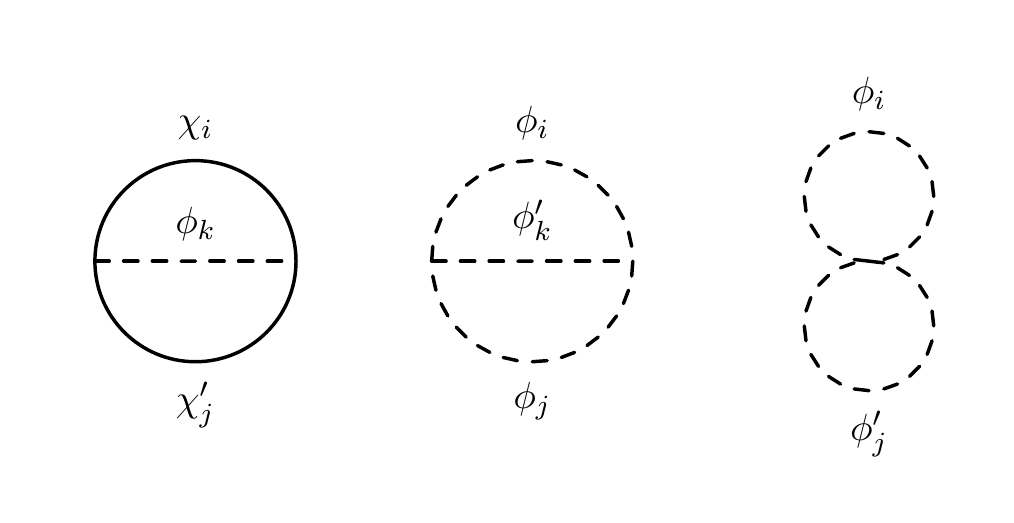}
\caption{Feynman diagrams contributing to $(\alpha_\lambda + \alpha_\kappa)^2$ corrections to the effective potential at the two-loop 
level. 
These consist of the corrections from neutralinos/charginos 
$\chi_i,\chi'_i \equiv \tilde{\chi}^0_i, \tilde{\chi}^+_i$,  and Higgs particles $\phi_i,\phi'_i \equiv  h_i,A_i^0,H_i^+$. 
A sum over all possible generations is assumed, but all CP and charge violating diagrams are not considered. }
\label{fig:DiagramsLambda}
\end{figure}

The calculation of $\Veff^{(2)}$ is performed in the gaugeless limit and includes all possible two-loop corrections except those 
involving the electroweak gauge couplings $g_1$ and $g_2$. 
We list the corresponding diagrams in Figs.~\ref{fig:DiagramsAlphaS}--\ref{fig:DiagramsLambda}. The corrections of 
$O(\alpha_S(\alpha_t+\alpha_b))$ stemming from the diagrams in Fig.~\ref{fig:DiagramsAlphaS} had already been calculated in the 
context of the NMSSM \cite{Degrassi:2009yq} and the perfect agreement between the results obtained by 
\SARAH/\SPheno and those of Ref.~\cite{Degrassi:2009yq}
was shown in Ref.~\cite{Goodsell:2014bna}. The classes of diagrams shown in Fig.~\ref{fig:DiagramsYukawa} have so far only been calculated in the 
MSSM \cite{Brignole:2001jy,Degrassi:2001yf,Brignole:2002bz,Dedes:2002dy,Dedes:2003km}
but not in the NMSSM. Finally, contributions similar to Fig.~\ref{fig:DiagramsLambda} are completely absent in the MSSM 
in the limit $g_1=g_2=0$ but are present in the NMSSM, so these represent a novel contribution.

\section{Numerical Results}
\label{sec:numerics}
\subsection{Numerical setup}
For the numerical analysis we generated with {\tt SARAH 4.4.0} the Fortran code for \SPheno 
which was compiled with {\tt SPheno 3.3.3}. We applied by hand a few changes to that code to 
turn on/off different two-loop contributions individually. In addition, we included for 
comparison the 
$\alpha_S (\alpha_t + \alpha_b)$ corrections from the NMSSM of Ref.~\cite{Degrassi:2009yq} as well as the 
MSSM corrections of Refs.~\cite{Brignole:2001jy,Degrassi:2001yf,Brignole:2002bz,Dedes:2002dy,Dedes:2003km} 
for all other contributions. All parameter scans have been performed with {\tt SSP} \cite{Staub:2011dp}.

\subsection{Heavy singlet case with moderate $\lambda$}
To test the importance of the two-loop corrections beyond $O(\alpha_s(\alpha_t+\alpha_b))$ we start with a 
parameter point in the constrained NMSSM. In this constrained setup, universal boundary conditions at the GUT scale 
are assumed:
\begin{eqnarray*}
& M_1 = M_2 = M_3 \equiv M_{1/2} & \\
& m_{\tilde{D}}^2 = m_{\tilde{U}}^2 = m_{\tilde{Q}}^2 = m_{\tilde{E}}^2 = m_{\tilde{L}}^2 \equiv m_0^2 \, {\bf 1}_3 & \\
& T_i \equiv A_0 Y_i \hspace{0.5cm} i=u,d,e; \hspace{1cm} T_\lambda \equiv A_\lambda \lambda; \quad \hspace{1cm} \quad T_\kappa \equiv A_\kappa \kappa & 
\end{eqnarray*}
$M_{1/2}$, $m_0$, $A_0$, $A_\lambda$ and $A_\kappa$ are defined at the unification scale, while $\lambda$, $\kappa$, $\mueff$ and $\tan\beta = \frac{v_u}{v_d}$ are defined at the SUSY scale. As an example 
for the discussion in the following we pick the point fixed by
\begin{eqnarray}
&m_0 = 1.4~\text{TeV} \,\hspace{0.5cm} 
M_{1/2} = 1.4~\text{TeV} \,\hspace{0.5cm} 
\tan\beta = 2.9 \,\hspace{0.5cm} 
A_0 = -1.35~\text{TeV} \,\hspace{0.5cm} \nonumber &\\
\label{eq:PointGUT}
&\lambda=0.56 \,\hspace{0.5cm}
\kappa =0.33  \,\hspace{0.5cm}
A_\lambda = -390~\text{GeV}  \,\hspace{0.5cm}
A_\kappa = -280~\text{GeV}  \,\hspace{0.5cm}
\mueff = 200~\text{GeV}&
\end{eqnarray}
The results for the Higgs masses for this point at different loop level and with different loop corrections are summarised 
in Tab~\ref{tab:CNMSSM}. Note that it is in this case the \emph{second} neutral scalar $h_2$ that is predominantly singlet in this case (about $96\%$); we find at two loops
$$\phi_s =0.09 h_1 - 0.98 h_2 + 0.17 h_3.$$
In this case the mass of the mostly singlet scalar is of the order of $200$ GeV, yet it does not mix substantially with the lightest Higgs state -- justifying this point as being labelled ``heavy''.

\begin{figure}[hbt]
\includegraphics[width=0.5\linewidth]{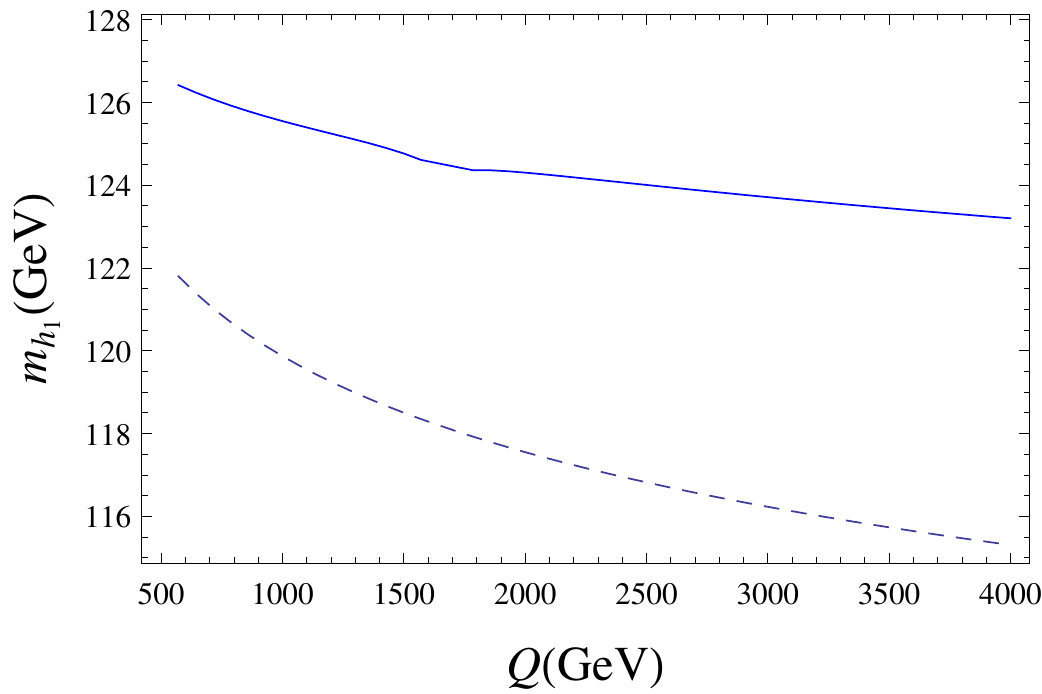} 
\caption{The light Higgs mass based on the parameter point of eq.~\ref{eq:PointGUT} for a variation of 
the renormalisation scale $Q$. The dashed line is the one-loop prediction and the full line the two-loop one.}
\label{fig:Q}
\end{figure}

\begin{table}[hbt]
\begin{tabular}{|c|cccc|}
\hline 
   & \parbox{0.15\linewidth}{Tree} & \parbox{0.15\linewidth}{one-loop} & \parbox{0.15\linewidth}{two-loop \\ ($\alpha_s(\alpha_b+\alpha_t)$)}  &  \parbox{0.15\linewidth}{full \\ two-loop}  \\
\hline 
$m_{h_1}$   &   93.8   &  117.6 (+ 25.4\%)   &  126.1 (+7.2\%)  &  124.7 (-1.1\%)   \\ 
$m_{h_2}$   &   214.5  &  209.2 (-2.4\%)     &  209.2 ($\pm$ 0\%) &  208.7 (-0.2\%)  \\
$m_{h_3}$   &   555.5  &  541.9 (-2.4\%)     &  542.3 (+0.1\%)  &  541.4 (-0.2\%)  \\
\hline
\end{tabular}
\caption{Higgs masses at tree-level, one-loop and two-loop for the parameter point of eq.~(\ref{eq:PointGUT}) in GeV. Note that the second Higgs $h_2$ is mostly singlet-like, with the MSSM-like heavy Higgs being the heaviest ($h_3$).}
\label{tab:CNMSSM}
\end{table}

At the two-loop level we distinguish two different cases: (i) including just $\alpha_s(\alpha_t + \alpha_b)$ which 
had been calculated before for the NMSSM; (ii) the full NMSSM corrections calculated by 
\SARAH/\SPheno. We see in Tab.~\ref{tab:CNMSSM} that the tendencies are similar to the MSSM: 
while the corrections proportional to the strong interaction give a large positive shift to the Higgs masses, the 
corrections involving only superpotential couplings are negative for all three Higgs masses. If one compares 
these numbers with the same corrections in the CMSSM one sees that the two-loop corrections are dominated 
for this point by MSSM-like contributions: for the CMSSM point with $m_0=M_{1/2}=1.4$~TeV, $\tan\beta=2.9$, $\mu>0$
and $A_0=-1.35$~TeV, the contributions involving the strong-interaction cause a shift by +11.3~GeV, while the purely
Yukawa corrections reduce the mass by -1.4~GeV.

To give an impression 
of the remaining uncertainty in the Higgs mass we show in Fig.~\ref{fig:Q} the calculated mass at one and 
two loops for a variation of the renormalisation scale $Q$. We see that the scale dependence is highly reduced at 
the two-loop level as expected. The average $M_{\text{SUSY}}=\sqrt{m_{\tilde{t}_1} m_{\tilde{t}_2}}$ of the stops (usually taken
as the optimal renormalisation scale) is about $2$~TeV. The variation of the Higgs mass in the range $\frac{1}{2} M_{\text{SUSY}}$ 
and $2 M_{\text{SUSY}}$ is $2.3$~GeV which can be taken as a rough indication of the remaining uncertainty. 

\begin{figure}[hbt]
\includegraphics[width=0.67\linewidth]{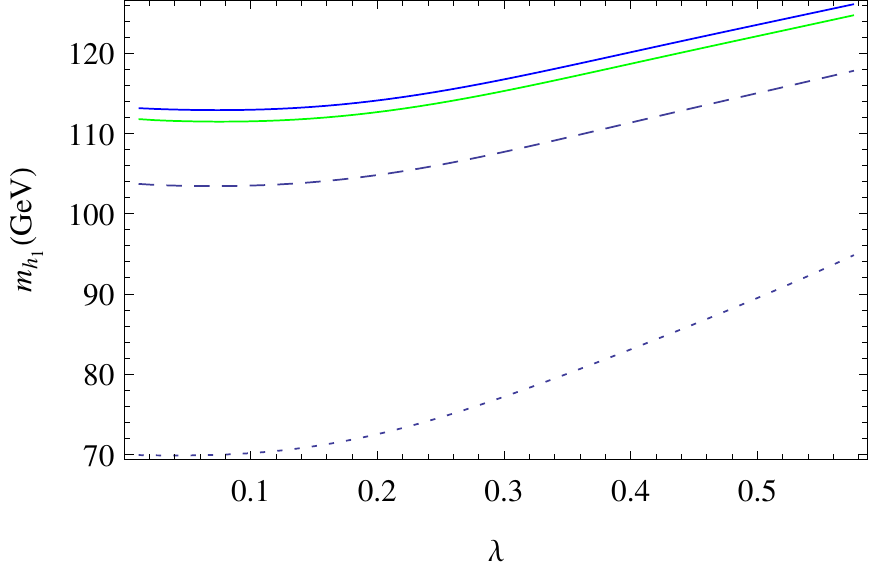}  \\
\includegraphics[width=0.67\linewidth]{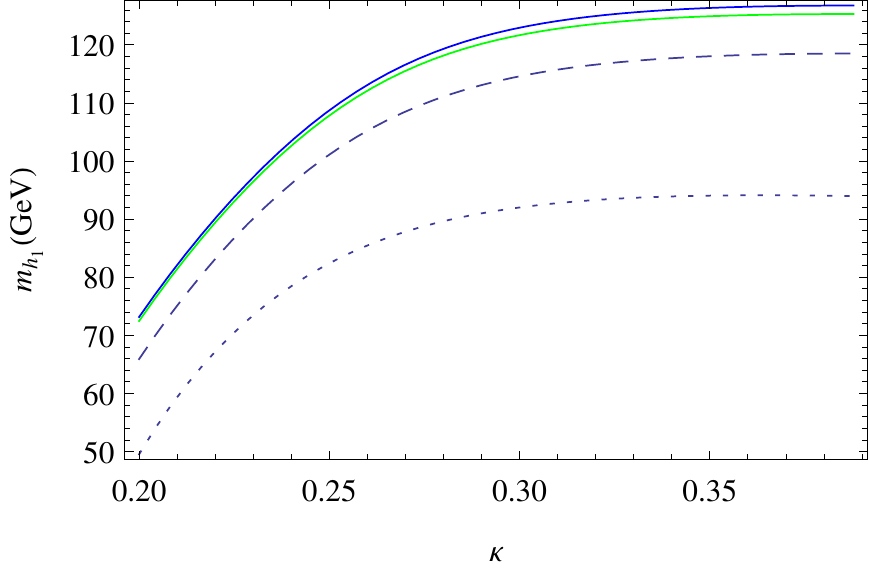}  
\caption{Light Higgs mass based on the parameter point of eq.~\ref{eq:PointGUT} for a variation of 
$\lambda$ and $\kappa$. The Higgs mass is shown at tree-level (dotted line), one-loop (dashed line) and two-loops (full line). 
At two-loop level we distinguish between the $\alpha_S(\alpha_b + \alpha_t)$ corrections (blue)
and the full calculation in the NMSSM (green). }
\label{fig:NMSSM_GUT}
\end{figure}

The general picture of the importance of the loop corrections is 
also confirmed when we vary $\lambda$ and $\kappa$ as shown in Fig.~\ref{fig:NMSSM_GUT}. 

\begin{figure}[hbt]
\includegraphics[width=0.4\linewidth]{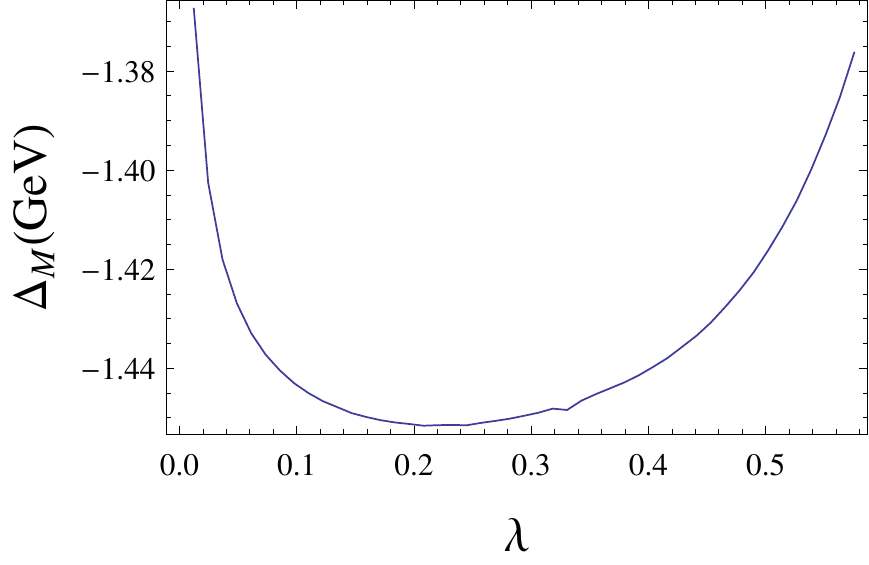}  \hspace{1cm}
\includegraphics[width=0.4\linewidth]{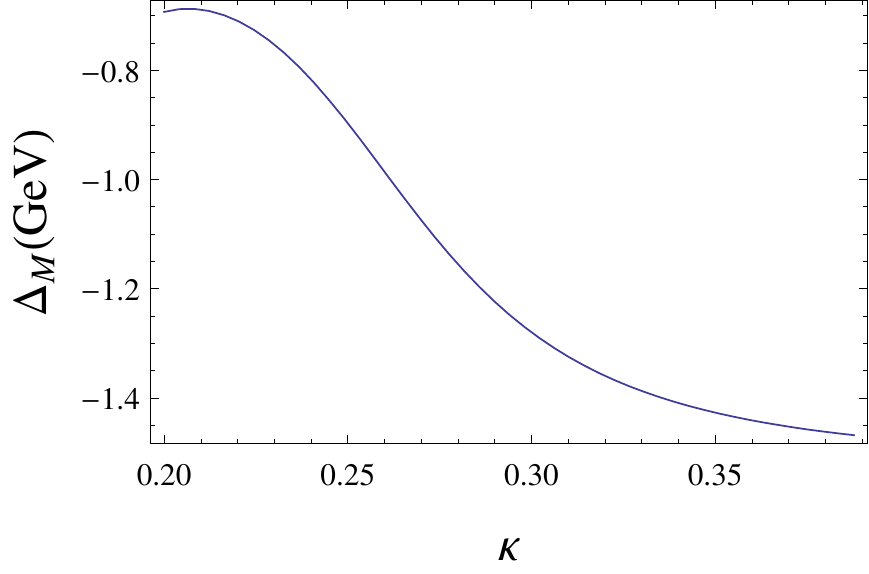}  \\
\includegraphics[width=0.4\linewidth]{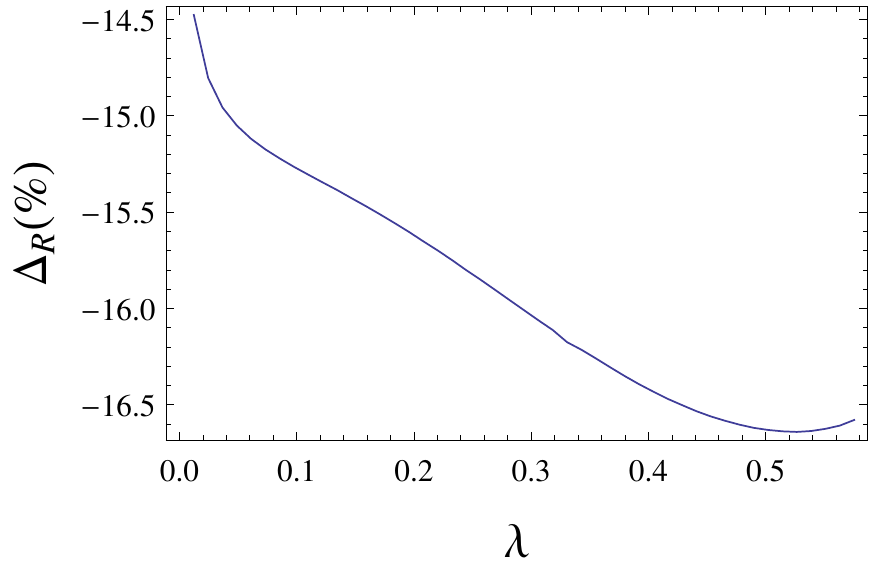}  \hspace{1cm}
\includegraphics[width=0.4\linewidth]{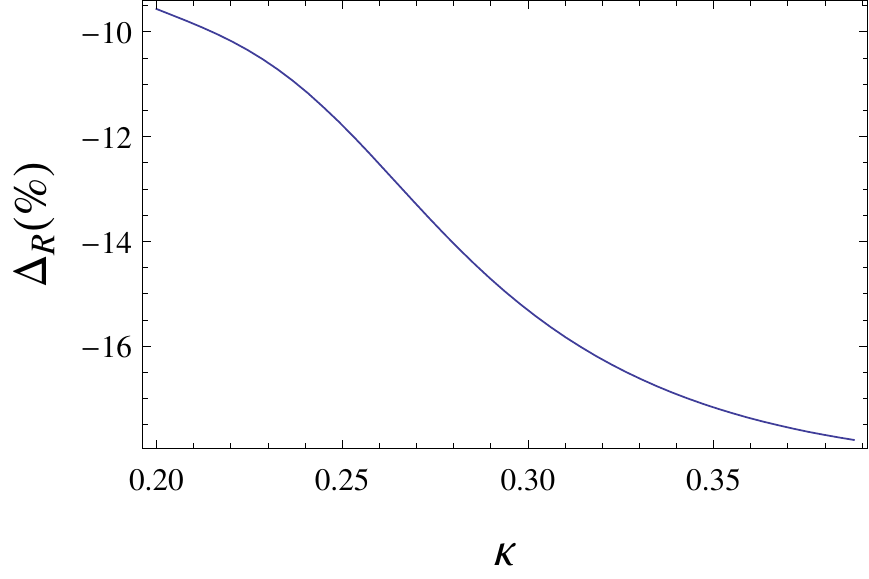}  
\caption{First row: the absolute size of 
the two-loop contributions beyond $O(\alpha_S(\alpha_b + \alpha_t))$ for the parameter variation shown 
in Fig.~\ref{fig:NMSSM_GUT}. Second row:  the 
relative size of these corrections normalized to the $\alpha_S(\alpha_b + \alpha_t)$ ones. }
\label{fig:NMSSM_GUT_rel}
\end{figure}
In general, the additional corrections are small compared to the corrections involving the strong interaction:
for the above parameters the size of these corrections is 10\%--20\% of the size of the $\alpha_s$ corrections and
dominated by the MSSM-like contributions. This also explains the very moderate dependence of the overall
size of these corrections on the NMSSM specific parameters $\lambda$ and $\kappa$. 

\subsection{Heavy singlet with large $\lambda$}
So far we concentrated on moderate values of $\lambda$, which are consistent with gauge coupling unification and do not 
cause a Landau pole below the unification scale. However, if one surrenders the condition of perturbativity up to the 
GUT scale larger values of $\lambda$ are possible. These so-called $\lambda$SUSY scenarios\footnote{$\lambda$SUSY as typically defined allows for more general soft-breaking and superpotential terms. Here we restrict to the large $\lambda$ limit of the NMSSM.} are popular because 
they predict very moderate values for the fine-tuning \cite{Hall:2011aa} and have interesting phenomenological 
consequences \cite{SchmidtHoberg:2012yy,SchmidtHoberg:2012ip}. We consider in the following the point fixed by
\begin{eqnarray}
& \lambda = 1.6 \,\hspace{0.5cm} 
\kappa = 1.6 \,\hspace{0.5cm}  
\tan\beta = 3 \,\hspace{0.5cm}  
T_\lambda = 600~\text{GeV}\,\hspace{0.5cm} 
T_\kappa = -2650~\text{GeV} \,\hspace{0.5cm} 
\mueff = 614~\text{GeV} \,\hspace{0.5cm} & 
\label{eq:lambdaSUSYpoint}
\end{eqnarray}
All sfermion squared soft-masses are fixed to $2\cdot 10^6~\text{GeV}^2$, the gaugino masses are $M_1=200$~GeV,
$M_2=400$~GeV, $M_3 = 2000$~GeV while all trilinear MSSM-like soft-terms are assumed to vanish. The corresponding masses 
are shown in Tab.~\ref{tab:LambdaSUSY}. Again we find that the MSSM-heavy-Higgs-like state is the heaviest of the three neutral scalars, with the mostly-singlet neutral scalar having a mass of order $700$ GeV. 
\begin{table}[hbt]
\begin{tabular}{|c|cccc|}
\hline 
   & \parbox{0.15\linewidth}{Tree} & \parbox{0.15\linewidth}{one-loop} & \parbox{0.15\linewidth}{two-loop \\ ($\alpha_s(\alpha_b+\alpha_t)$)}  &  \parbox{0.15\linewidth}{full \\ two-loop}  \\
\hline 
$m_{h_1}$   &   144.8   &  122.6 (-15.3\%)  &  126.5 (+3.2\%)  &    128.0 (+1.2\%)     \\ 
$m_{h_2}$   &   713.2   &  745.9 (+4.6\%)  &  745.8 (+0.0\%)  &     747.9  (+0.3\%)   \\
$m_{h_3}$   &   1454.5  &  1421.1  (-2.3\%) &  1420.1 (-0.1\%)  &   1420.3 (+0.0\%)   \\
\hline
\end{tabular}
\caption{Higgs masses at tree-level, one-loop and two-loop for the parameter point of eq.~(\ref{eq:lambdaSUSYpoint}). Again the \emph{second} Higgs $h_2$ is mostly singlet-like, with the heavy Higgs $h_3$ being largely the MSSM heavy Higgs.}
\label{tab:LambdaSUSY}
\end{table}
\begin{figure}[hbt]
\includegraphics[width=0.5\linewidth]{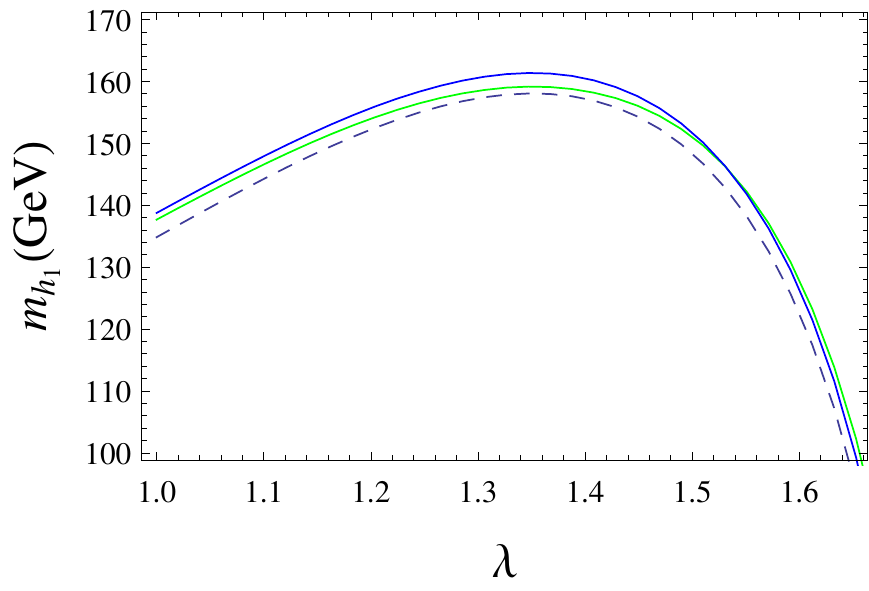} 
\caption{The light Higgs mass based on the parameter point of eq.~\ref{eq:lambdaSUSYpoint} for a variation of 
$\lambda$. The colour and line coding is the same as in Fig.~\ref{fig:NMSSM_GUT}}
\label{fig:NMSSM_lambdaSUSY_lam}
\end{figure}
We see that for this point all two-loop shifts to the lightest Higgs mass have the same sign, thus the NMSSM-specific corrections 
at some point dominate the ones of the MSSM. This is depicted in Fig.~\ref{fig:NMSSM_lambdaSUSY_lam} where the light 
Higgs mass for a variation of $\lambda$ is shown: for $\lambda \simeq 1.5$ the sign of the corrections beyond $\alpha_S(\alpha_t+\alpha_b)$
changes and those contributions become positive. This is depicted again on the left of Fig.~\ref{fig:NMSSM_lambdaSUSY_lam_zoom}
where the absolute size of the $\alpha_S(\alpha_t+\alpha_b)$ and $(\alpha_\lambda+\alpha_\kappa+\alpha_t+\alpha_b+\alpha_\tau)^2$
corrections are shown. On the right of Fig.~\ref{fig:NMSSM_lambdaSUSY_lam_zoom} we zoom into the interesting range with a SM-like
Higgs mass in the preferred range. In this plot we also  show the expected mass if only MSSM-like corrections for 
the purely Yukawa part are included. One sees that this can give masses which are wrong by several GeV.

\begin{figure}[hbt]
\includegraphics[width=0.49\linewidth]{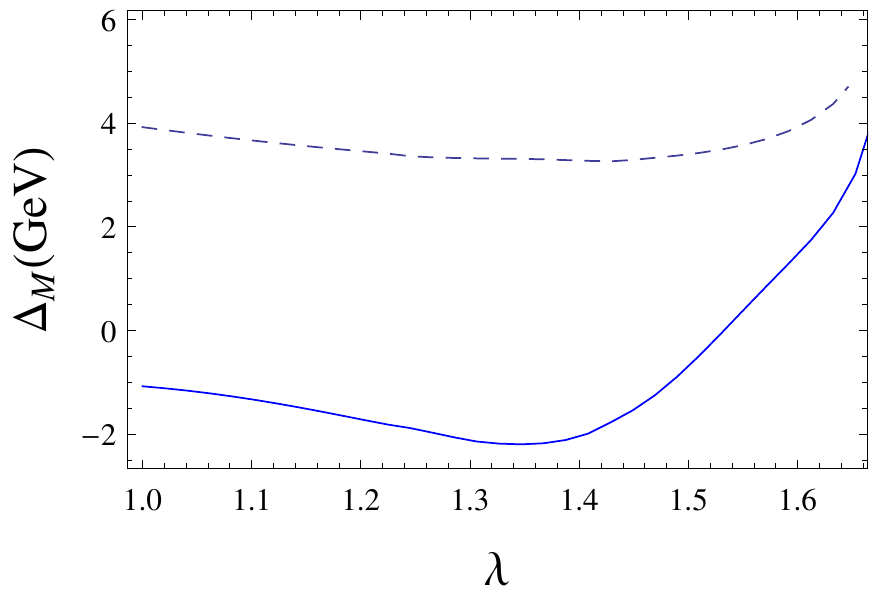} \hfill
\includegraphics[width=0.49\linewidth]{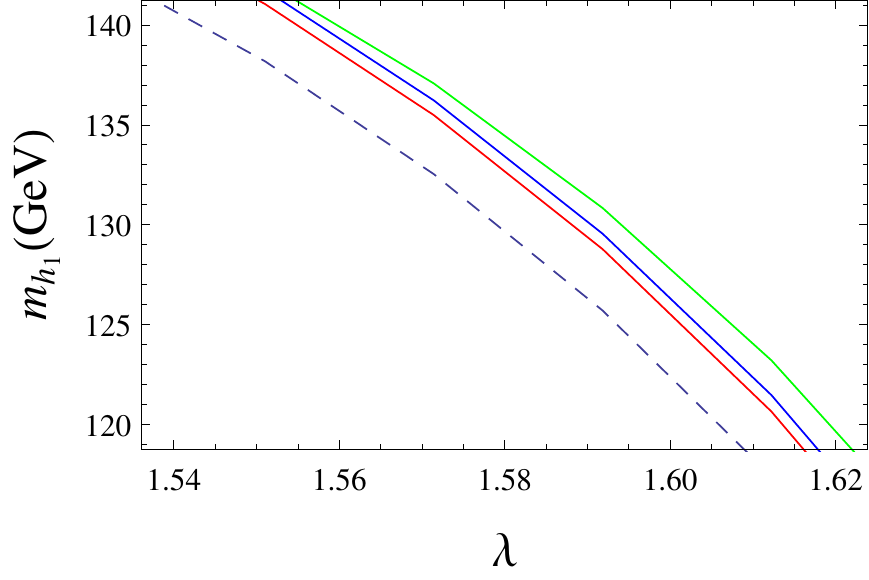} \hfill 
\caption{On the left: absolute size of the two-loop corrections $O(\alpha_S(\alpha_t+\alpha_b))$ (dashed) 
and $O((\alpha_\lambda+\alpha_\kappa+\alpha_t+\alpha_b+\alpha_\tau)^2)$ (full line). On 
the right: zoom into the interesting region of Fig.~\ref{fig:NMSSM_lambdaSUSY_lam} with 
a SM-like Higgs state in the preferred range. The blue line is the mass including $\alpha_S(\alpha_b+\alpha_t)$ corrections, 
the green line is the full calculation while the red line corresponds to MSSM-like corrections only for the Yukawa part.  }
\label{fig:NMSSM_lambdaSUSY_lam_zoom}
\end{figure}

\subsection{Light singlet case}
We have seen that in the case of a heavy singlet the additional 
two-loop contributions to the ones involving the strong interaction 
are MSSM-like. We discuss now the effects if a light singlet 
is present. For this purpose we consider the benchmark point BMP-A of Ref.\cite{Das:2014fha} which has the interesting feature 
that all three scalars have masses below 200~GeV. The important parameter values are 
\begin{eqnarray}
&\lambda=0.596 \,\hspace{0.5cm}
\kappa =0.596  \,\hspace{0.5cm}
T_\lambda = -27\ \text{GeV}  \,\hspace{0.5cm}
T_\kappa = -240~\text{GeV}  \,\hspace{0.5cm}
\mueff = 130~\text{GeV}& \nonumber \\
&T_t = -3050~\text{GeV}\,\hspace{0.5cm}
T_b = T_\tau = -1000~\text{GeV}\,\hspace{0.5cm}
m^2_{\tilde{Q},33} = 9.0\cdot 10^5~\text{GeV}^2\,\hspace{0.5cm}
m^2_{\tilde{U},33} = 1.05\cdot 10^6~\text{GeV}^2& \nonumber \\
\label{eq:PointLightScalars}
\end{eqnarray}
We give in Tab.\ref{tab:NMSSMlight} the values for the three scalar masses at the different loop levels. One sees
here that in general the loop corrections to the SM-like Higgs, which is the second mass eigenstate, are less 
important than in the case of a heavy singlet because of the already enhanced tree-level mass. However, the 
interesting feature is that the one- and two-loop corrections can be of similar size and the most important 
corrections are the two-loop parts discussed here for the first time. 

\begin{table}[hbt]
\begin{tabular}{|c|cccc|}
\hline 
   & \parbox{0.15\linewidth}{Tree} & \parbox{0.15\linewidth}{one-loop} & \parbox{0.15\linewidth}{two-loop \\ ($\alpha_s(\alpha_b+\alpha_t)$)}  &  \parbox{0.15\linewidth}{full \\ two-loop}  \\
\hline 
$m_{h_1}$   &  19.4    &  67.8 (+249.5\%)    &  74.5 (+9.9\%)  &   74.2 (-0.4\%)    \\ 
$m_{h_2}$   &  122.7   &  123.5 (+0.7\%)   &  124.3 (+0.6\%) &   123.3  (-0.8\%)  \\
$m_{h_3}$   &  177.4   &  188.2 (+6.1\%)  &  192.7  (+2.3\%) &   191.1  (-0.8\%) \\
\hline
\end{tabular}
\caption{Higgs masses at tree-level, one-loop and two-loop for the parameter point of eq.~(\ref{eq:PointLightScalars}). }
\label{tab:NMSSMlight}
\end{table}

\begin{figure}
\includegraphics[width=0.67\linewidth]{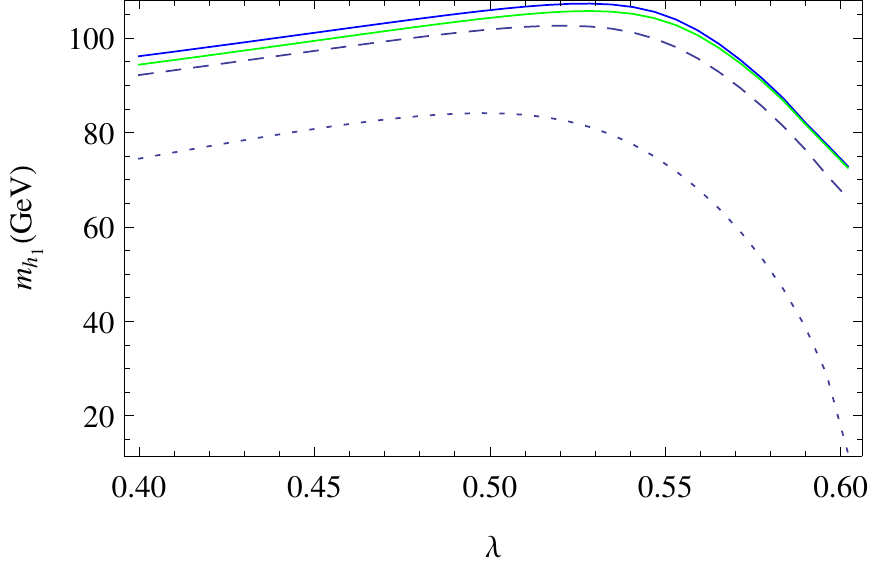} \\
\includegraphics[width=0.67\linewidth]{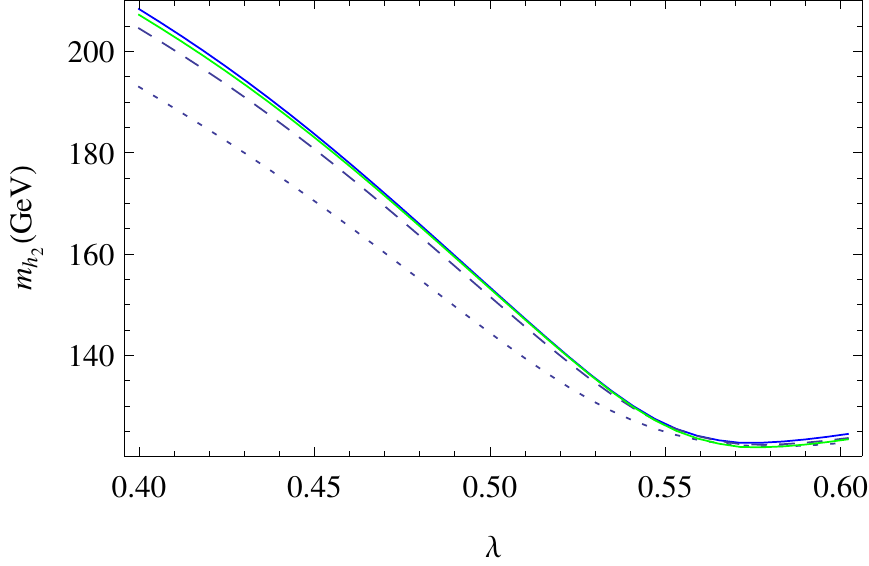} 
\caption{The two lightest Higgs masses based on the parameter point of eq.~\ref{eq:PointLightScalars} for a variation 
of $\lambda$. The colour code is the same as in Fig.~\ref{fig:NMSSM_GUT}}
\label{fig:NMSSM_LightSinglet_Masses}
\end{figure}

\begin{figure}[hbt]
\includegraphics[width=0.49\linewidth]{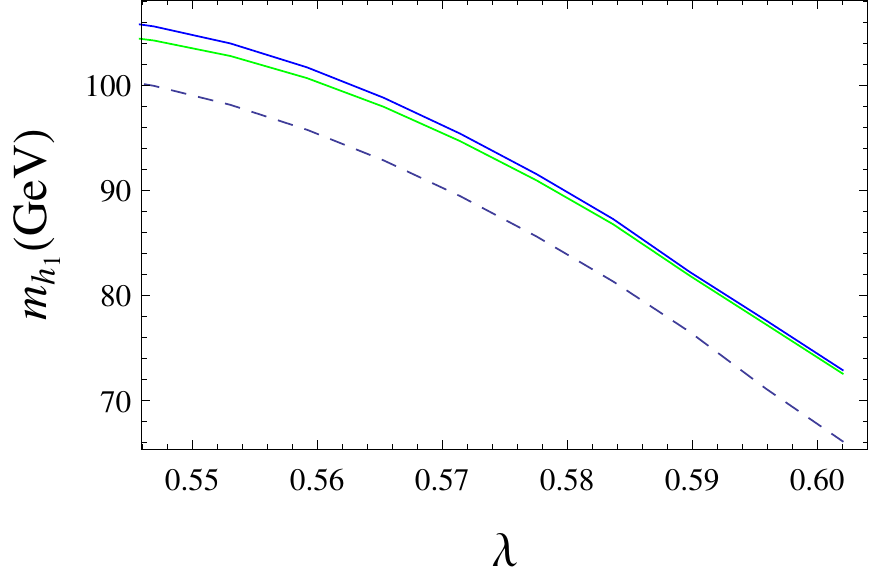} \hfill
\includegraphics[width=0.49\linewidth]{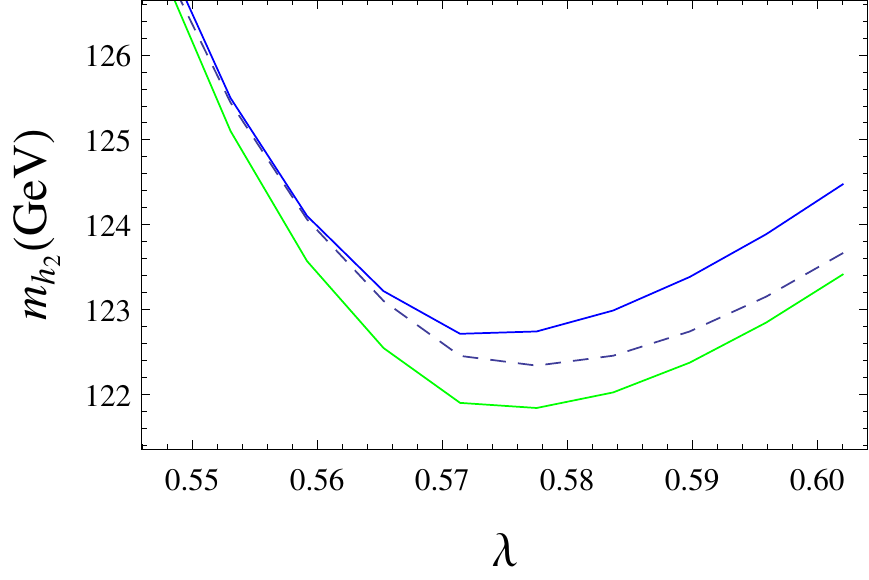} \\
\includegraphics[width=0.49\linewidth]{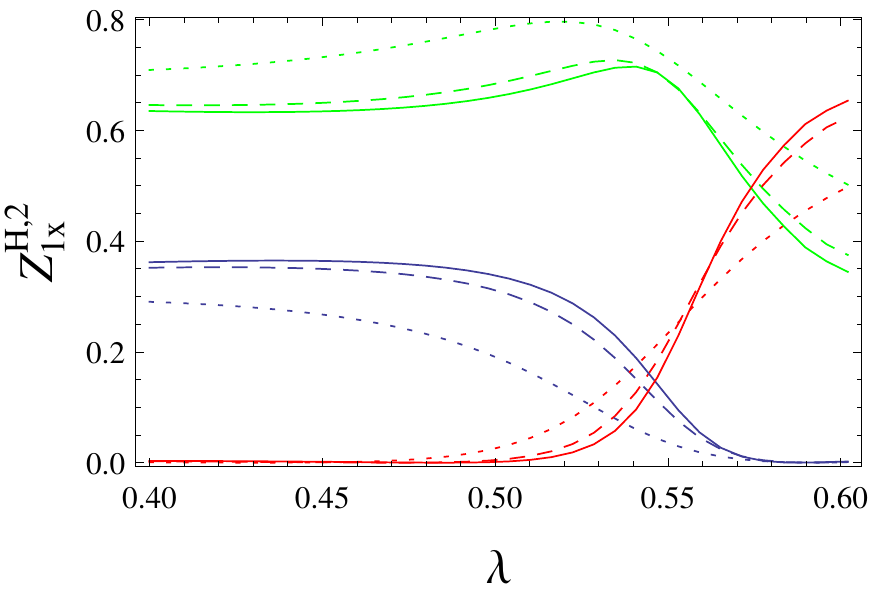} \hfill
\includegraphics[width=0.49\linewidth]{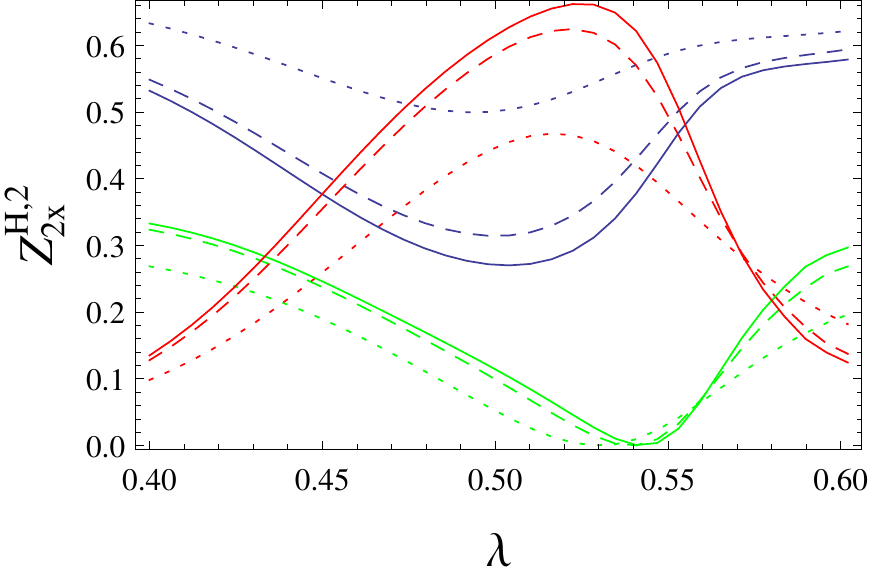} \\
\includegraphics[width=0.49\linewidth]{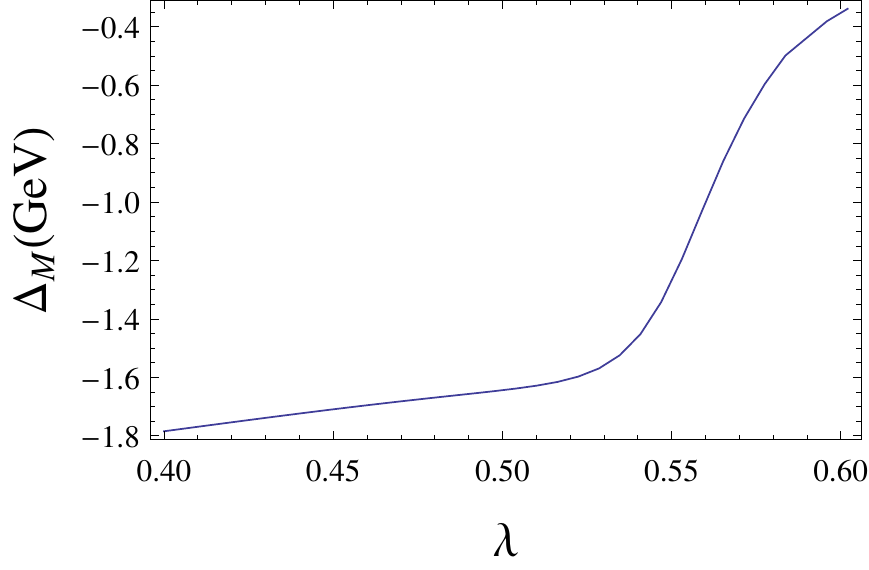} \hfill
\includegraphics[width=0.49\linewidth]{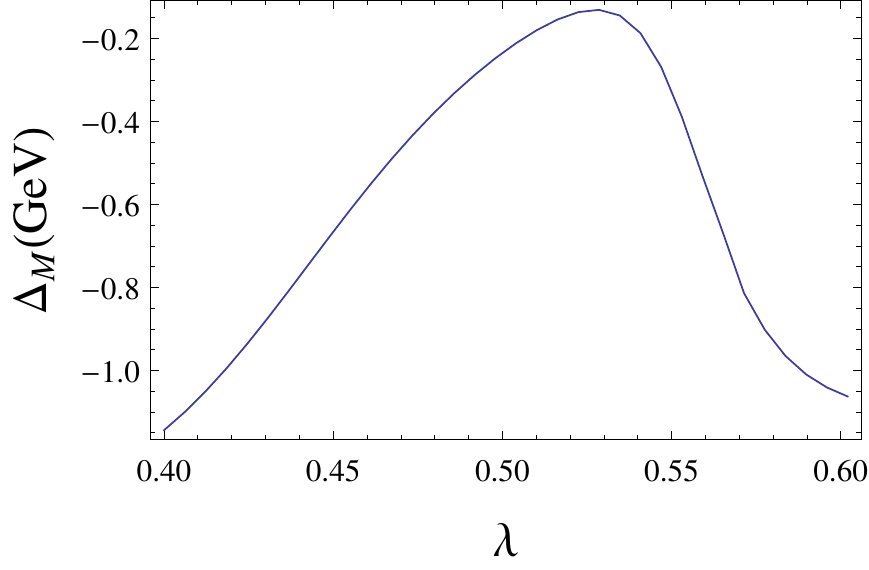} 
\caption{First row: zoom into the range $\lambda \in [0.4,0.6]$ of Fig.~\ref{fig:NMSSM_LightSinglet_Masses}.
The colour and line coding is the same as in  Fig.~\ref{fig:NMSSM_GUT}.
Second row: down- (blue), up- (green) and singlet- (red) fraction  of the Higgs particle at tree-level (dotted), one-loop (dashed) 
and two-loop (full line). Third row: absolute size of the two-loop corrections not involving the strong interaction, for $m_{h_1}$ on the left and $m_{h_2}$ on the right.}
\label{fig:NMSSM_LightSinglet_Masses_Zoom}
\end{figure}

To find a better understanding of the different effect, we depict 
the masses at tree-, one-loop and two-loop level for a variation of $\lambda$ in Fig.~\ref{fig:NMSSM_LightSinglet_Masses}.
One can see a very strong dependence of both masses on $\lambda$ which is however mainly caused by tree-level effects as 
indicated by the dotted line in Fig.~\ref{fig:NMSSM_LightSinglet_Masses}: for small $\lambda$ the singlet is heavier than 
the doublet which causes a reduction of the tree-level mass due to mixing effects and the SM-like state is much too light. 
With increasing $\lambda$ the singlet becomes lighter and for $\lambda \simeq 0.55$ a level crossing takes place. The 
mixing with the lighter singlet as well as $F$-term contributions $\delta m_h \sim \frac{\lambda^2 v^2}{2} \cos^2 2\beta$ 
give a sizeable push to the SM-like Higgs state at tree-level. Thus, the interesting region is the one close to and above 
the level crossing and we zoom into this region in Fig.~\ref{fig:NMSSM_LightSinglet_Masses_Zoom}. 

We give in this Figure also the composition of the two light Higgs states as well as the absolute size of the full two-loop corrections 
compared to the $\alpha_S(\alpha_t + \alpha_b)$ ones. One sees a strong correlation between the singlet fraction and the 
two-loop corrections not involving the strong interaction shown in the second and third row of Fig.~\ref{fig:NMSSM_LightSinglet_Masses_Zoom}:
the absolute size of these corrections decreases with increasing singlet admixture. Thus, the main contribution to the two-loop 
masses despite the sizeable value of $\lambda$ is again caused by the MSSM-like corrections due to (s)quarks. 
Interestingly, for a light state below 80~GeV which is 60\% singlet, the $\alpha_S(\alpha_b+\alpha_t)$ can still give a sizeable push 
while the additional corrections nearly vanish. In contrast, for the SM-like particle and $\lambda > 0.55$ the additional corrections 
discussed here can even dominate the strong corrections. Thus, in these cases the incomplete calculation would give the wrong picture that 
the two-loop corrections increase the Higgs mass. 

\begin{figure}
\includegraphics[width=0.49\linewidth]{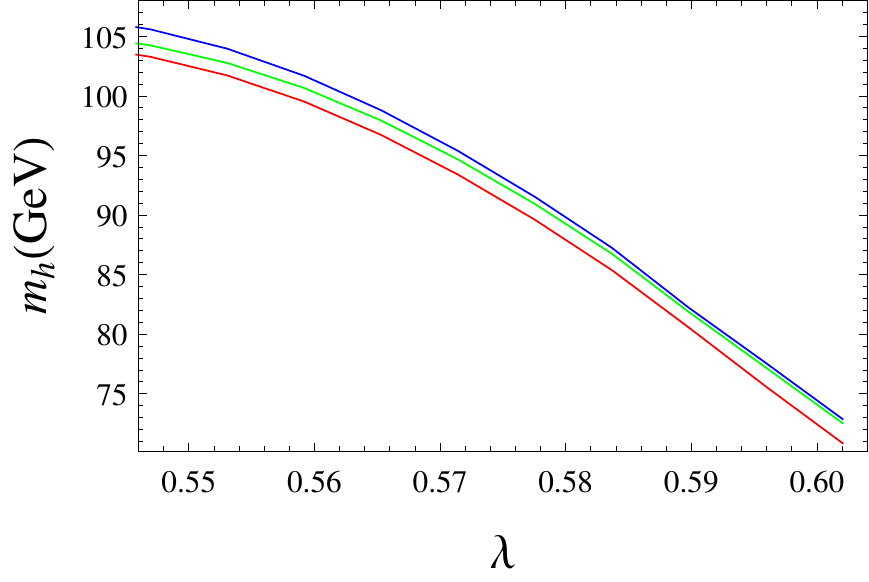} \hfill
\includegraphics[width=0.49\linewidth]{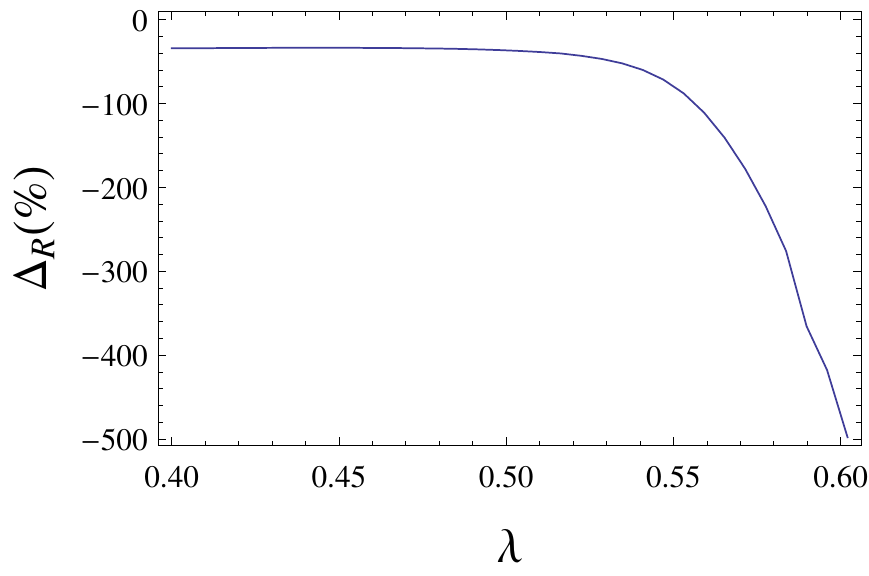} 
\caption{On the left: the lightest Higgs mass at two loops based on the parameter point of eq.~\ref{eq:PointLightScalars} for a variation 
of $\lambda$. The blue line is the mass using only $\alpha_s(\alpha_t+\alpha_b)$ corrections, the 
green line corresponds to our full calculation while the red line gives the obtained result when using the MSSM expressions for the 
pure Yukawa contributions. On the right we show the relative difference between the full corrections and the MSSM approximation. }
\label{fig:MSSMapprox}
\end{figure}

Finally, we comment again on the approximation of using the known two-loop Yukawa corrections of the MSSM to the upper $2\times 2$ block of the scalar mass matrix. While, as we have shown above, this can sometimes be a good approximation for a MSSM-like state for moderate $\lambda$, we found in the previous subsections 
that this fails for larger $\lambda$. Another situation where this approximation gives wrong results 
is the case of light singlets as shown in Fig.~\ref{fig:MSSMapprox}. 

We see in this Figure that the approximation to use the MSSM results for the purely Yukawa interactions roughly works for smaller values of $\lambda$ 
but becomes worse for increasing $\lambda$. For $\lambda \simeq 0.6$ this approximation predicts a change of about 2~GeV of the light Higgs mass 
compared to the full calculation of the NMSSM. This effect is mainly caused by the missing corrections to the $(1,3)$ and $(2,3)$ 
elements in the Higgs mass matrix. Since the corrections are negative for these entries and reduce the mixing between the doublets and singlet, the 
incomplete calculation predicts a mixing that is too large, which reduces the lighter mass eigenstate. 

\section{Discussion}
 \label{sec:conclusion}

We have examined the impact of the dominant two-loop corrections beyond $O(\alpha_S(\alpha_t+\alpha_b))$ in the NMSSM. The previously best approach was to use the $O(\alpha_S(\alpha_t+\alpha_b))$ routines for the NMSSM of \cite{Degrassi:2009yq} and add the MSSM contributions for $O((\alpha_t + \alpha_b + \alpha_\tau)^2)$. We have found, as could be expected, that for small $\lambda$ this is a good approximation. It should also not be a surprise that it is a poor approximation when there is significant mixing between the singlet and the MSSM-like states, with an error of approximately $2$~GeV. 
On the other hand, we found that the approximation works up to somewhat large values of $\lambda$ when the singlet was heavier than the standard-model-like Higgs, although for $\lambda$ above around $1.5$ the full contribution has a different sign to the previous best approximation and so the correct result differs by a few GeV. To examine this more closely, let us consider the one-loop approximation for the Higgs mass including the stop corrections and leading $\lambda^4$ contributions (from singlet scalars and Higgsinos) in the decoupling limit of the effective potential approximation:
\begin{align}
m_h^2 \approx& M_Z^2 c_\beta^2 + \frac{\lambda^2 v^2s_{2\beta}^2}{2} + \frac{3}{2\pi^2} \frac{m_t^4}{v^2} \log \frac{m_{\tilde{t}_1} m_{\tilde{t}_2}}{m_t^2} + \frac{\lambda^4 v^2}{32\pi^2} \log \frac{m_{\phi_s\phi_s}^2m_{\sigma_s\sigma_s}^2}{(4\kappa^2 \mueff^2/\lambda^2)^2}.
\end{align}
 We see at one loop:
\begin{enumerate}
\item As is well-known, for small $\tan \beta$ the tree-level mass is larger than in the MSSM, reducing the impact of radiative corrections.
\item The contributions from the singlet will typically be smaller than those of the stops unless we have a large $\lambda$, or $m_{\phi_s\phi_s}$ much larger than the stop mass.  
\item Moreover, to obtain a substantial contribution from the singlets requires a large splitting between the singlet mass and the Higgsino mass $\mueff$. 
\end{enumerate}
 Now, in the NMSSM (respecting a $\mathbb{Z}_3$ symmetry) it is difficult to obtain a large value for the physical singlet mass, and moreover to split it from the Higgsinos. To see this, consider the large $v_s$ limit of the potential, where
\begin{align}
V \simeq& \frac{1}{2} m_S^2 v_s^2 + \frac{1}{3\sqrt{2}} \kappa A_\kappa v_s^3 + \frac{1}{4} \kappa^2 v_s^4.
\end{align}
This yields
\begin{align}
m_{\phi_s\phi_s}^2 \simeq& \frac{\kappa}{\lambda} A_\kappa \mueff + 4 \mueff^2 \left(\frac{\kappa}{\lambda}\right)^2,\qquad m_{\sigma_s \sigma_s}^2 \simeq - 3 \frac{\kappa}{\lambda} A_\kappa \mueff.
\end{align}
However, 
\begin{align}
\mueff \simeq& \frac{\lambda}{4\kappa} \big( - A_\kappa + \sqrt{A_\kappa^2 - 8 m_S^2} \big).
\end{align}
We find that for a stable vacuum we require $A_\kappa^2 \gtrsim 9 m_S^2 $ and therefore $\mueff \sim -\frac{\lambda}{4\kappa} A_\kappa$, giving
\begin{align}
m_{\phi_s\phi_s}^2 \simeq& 4 \left( \frac{\kappa^2}{\lambda^2}\right) \mueff \sqrt{A_\kappa^2 - 8 m_S^2} < 4 \left( \frac{\kappa^2}{\lambda^2}\right) \mueff^2,\qquad m_{\sigma_s \sigma_s}^2 \simeq 12 \frac{\kappa^2}{\lambda^2}\mueff^2.
\end{align}
Hence to obtain a large hierarchy between the singlets and the singlino is not possible at tree level. There is also a similar contribution proportional to $v^2 \lambda^4 $ from the heavy Higgs and the higgsinos, which then requires a hierarchy between the heavy higgs and $\mueff$, which is similarly constrained. 
We therefore expect these observations for the heavy singlet limit to apply roughly at two loops, because the supersymmetric mass for the singlet is of the same order as the supersymmetry-breaking contribution. 

In summary, we have examined the full two-loop corrections to the Higgs masses (calculated in the ``gaugeless limit'' where gauge couplings of broken gauge groups are ignored) in the NMSSM and compared them to the approximations previously available. The new corrections implemented in \SARAH/\SPheno are essential to obtain an accurate calculation of the Higgs mass in the NMSSM when the values of $\lambda$ are significant and/or there is substantial mixing between the singlet scalar and the MSSM-like Higgses. It would be interesting, however, to explore other singlet extensions where the quantum effects could be even more significant -- for example the general NMSSM with a large $\lambda$, which would allow a genuinely large singlet scalar mass with a splitting between the scalar and Higgsinos.  This can be done now by an interested reader since the codes \SARAH and \SPheno are publically available at {\tt HepForge} ({\tt http://sarah.hepforge.org/}, {\tt http://spheno.hepforge.org/}) and \SARAH includes 
several additional singlet extensions in addition to the NMSSM.

\section*{Acknowledgments}

We would like to thank P.~Slavich for many helpful and illuminating discussions. 

\appendix

\section{Higgs sector of the NMSSM}

\subsection{Minimum Conditions of the Vacuum}

At tree level, the tadpole equations (the minimum conditions for the vacuum) are given by
\begin{equation}
	T_i = \frac{\partial V}{\partial v_i}\Big|_{\phi_i=0,\sigma_i=0} = 0 
	\label{EqTadPole}
\end{equation}
with
\begin{align} 
\label{eq:tadpoleD}
\frac{\partial V}{\partial \phi_{d}} &= m_{H_d}^2 v_d + \frac{1}{8} \Big(g_{1}^{2} + g_{2}^{2}\Big)v_d \Big(v_d^2  - v_u^2\Big) +\frac{1}{2} \Big( v_d \Big(v_{s}^{2} + v_{u}^{2}\Big)|\lambda|^2  -  v_{s}^{2} v_u \Re(\kappa\lambda^*)  -  \sqrt{2} v_s v_u  {\Re\Big(T_{\lambda}\Big)}\Big)\\ \label{eq:tadpoleU}
\frac{\partial V}{\partial \phi_{u}} &= m_{H_u}^2 v_u+\frac{1}{8} \Big(g_{1}^{2} + g_{2}^{2}\Big)v_u \Big(v_{u}^{2}- v_{d}^{2} \Big)
 +\frac{1}{2} \Big( \Big(v_{d}^{2} + v_{s}^{2}\Big)v_u |\lambda|^2  -  v_d v_{s}^{2} \Re(\kappa\lambda^*) -  \sqrt{2} v_d v_s  {\Re\Big(T_{\lambda}\Big)} \Big)\\ 
\label{eq:tadpoleS}
 \frac{\partial V}{\partial \phi_s} &=  m_S^2 v_s- v_d v_s v_u \Re(\lambda \kappa^*) +   v_{s}^{3} |\kappa|^2  + \frac{1}{2} \Big(v_s  ( v_{d}^{2} + v_{u}^{2} )|\lambda|^2 +\sqrt{2} \Big(- v_d v_u \Re(T_{\lambda}) + v_{s}^{2} \Re(T_{\kappa})\Big)\Big)
\end{align} 
All parameters in these equations are taken to be 
running $\overline{\mathrm{DR}}'$ \cite{Jack:1994rk} parameters at the renormalisation scale $Q$. 
In this context the VEVs $v_d$ and $v_u$ are derived from the running $Z$-boson mass
$m_Z(Q)$ of the $Z$-boson.  For more details of this approach we refer to Ref.~\cite{Pierce:1996zz}. 

The tadpole equations receive finite corrections at loop level. We calculate the one- 
and two-loop shifts denoted as $\delta t^{(1)}_i$ and $\delta t^{(2)}_i$, and demand that the sum of all orders vanishes
\begin{equation}
 T_i + \delta t^{(1)}_i  + \delta t^{(2)}_i= 0 \qquad{\rm for}\quad i=d,u,s .
\label{eq:oneloop}
\end{equation} 
The calculation of the corrections at one-loop 
is discussed in detail in Ref.~\cite{Staub:2010ty}. Details about the 
two-loop contributions are given in the text in sec.~\ref{sec:twoloopcalc}.

In our analysis we solve eqs.~(\ref{eq:oneloop}) for the soft SUSY breaking masses of the 
Higgs doublets and the singlet: $m^2_{H_d}, m^2_{H_u}$, and $m^2_{S}$. 

\subsection{Masses of the Higgs bosons}

\subsubsection{Tree-level scalar masses}
The tree-level mass matrices for the CP-even  Higgs bosons bosons
are calculated from the scalar potential via
\begin{equation}
 	m^{2,h}_{T,i j} = \frac{\partial^2 V}{\partial \phi_i \partial \phi_j}
 	   \Bigg|_{\phi_k=0,\sigma_k=0}, \hspace{2cm}
\label{ScalarMass}
\end{equation}
 with $i,j=1,2,3=u,d,s$. The matrix is symmetric 
and the entries  read
\begin{align} 
m_{\phi_{d}\phi_{d}} &= \frac{1}{2} \Big(v_{s}^{2} + v_{u}^{2}\Big)|\lambda|^2  + \frac{1}{8} \Big(g_{1}^{2} + g_{2}^{2}\Big)\Big(3 v_{d}^{2}  - v_{u}^{2} \Big) + m_{H_d}^2\nonumber\\ 
m_{\phi_{d}\phi_{u}} &= \frac{1}{4} \Big(-2 \sqrt{2} v_s {\Re\Big(T_{\lambda}\Big)}  + \Big(4 v_d v_u \lambda  - v_{s}^{2} \kappa \Big)\lambda^*  - v_{s}^{2} \lambda \kappa^* \Big) -\frac{1}{4} \Big(g_{1}^{2} + g_{2}^{2}\Big)v_d v_u \nonumber\\ 
m_{\phi_{u}\phi_{u}} &= \frac{1}{2} \Big(v_{d}^{2} + v_{s}^{2}\Big)|\lambda|^2  -\frac{1}{8} \Big(g_{1}^{2} + g_{2}^{2}\Big)\Big(-3 v_{u}^{2}  + v_{d}^{2}\Big) + m_{H_u}^2\nonumber\\ 
m_{\phi_{d}\phi_s} &= - \frac{1}{\sqrt{2}} v_u {\Re\Big(T_{\lambda}\Big)}  + v_s \Big(\Big(-\frac{1}{2} v_u \kappa  + v_d \lambda \Big)\lambda^*  -\frac{1}{2} v_u \lambda \kappa^* \Big)\nonumber\\ 
m_{\phi_{u}\phi_s} &= \frac{1}{2} \Big(- v_d \Big(\sqrt{2} {\Re\Big(T_{\lambda}\Big)}  + v_s \lambda \kappa^* \Big) - v_s \Big(-2 v_u \lambda  + v_d \kappa \Big)\lambda^* \Big)\nonumber\\ 
m_{\phi_s\phi_s} &= \frac{1}{2} \Big(2 \sqrt{2} v_s {\Re\Big(T_{\kappa}\Big)}  + \Big(6 v_{s}^{2} \kappa  - v_d v_u \lambda \Big)\kappa^*  + \Big(\Big(v_{d}^{2} + v_{u}^{2}\Big)\lambda  - v_d v_u \kappa \Big)\lambda^* \Big) + m_S^2 \label{eq:treescalarmatrix}
\end{align} 
This matrix is diagonalised by a rotation matrix $Z^H$ and the relation 
\begin{equation}
Z^{H} m^{2,h} Z^{H,T} = m^{2,h}_{\mathrm{diag}} \,.
\end{equation}
holds. The three eigenvalues of \(m^{2,h}_T\) correspond to
three physical mass eigenstates with masses $m_{h_1}$, 
$m_{h_2}$, $m_{h_3}$ which are ordered by their mass. 

\subsubsection{Tree-level pseudo-scalar masses}
\label{app:pseudoscalars}
Similarly, the mass matrix of the CP even states is calculated via
\begin{equation}
 	m^{2,A}_{T,i j} = \frac{\partial^2 V}{\partial \sigma_i \partial \sigma_j}
 	   \Bigg|_{\phi_k=0,\sigma_k=0}, \hspace{2cm}
\label{PseudoScalarMass}
\end{equation}
and one finds
\begin{align} 
m_{\sigma_{d}\sigma_{d}} &= \frac{1}{2} \Big(v_{s}^{2} + v_{u}^{2}\Big)|\lambda|^2  + \frac{1}{8} \Big(g_{1}^{2} + g_{2}^{2}\Big)\Big(- v_{u}^{2}  + v_{d}^{2}\Big) + m_{H_d}^2 + m_{\xi,\sigma_{d}\sigma_{d}}\nonumber\\ 
m_{\sigma_{d}\sigma_{u}} &=\frac{1}{4} v_s \Big(2 \sqrt{2} {\Re\Big(T_{\lambda}\Big)}  + 2 v_s {\Re\Big(\lambda \kappa^* \Big)} \Big) + m_{\xi,\sigma_{d}\sigma_{u}} \nonumber\\
m_{\sigma_{u}\sigma_{u}} &= \frac{1}{2} \Big(v_{d}^{2} + v_{s}^{2}\Big)|\lambda|^2  -\frac{1}{8} \Big(g_{1}^{2} + g_{2}^{2}\Big)\Big(- v_{u}^{2}  + v_{d}^{2}\Big) + m_{H_u}^2 + m_{\xi,\sigma_{u}\sigma_{u}\nonumber}\\ 
m_{\sigma_{d}\sigma_s} &= -\frac{1}{2} v_u \Big(2 v_s {\Re\Big(\lambda \kappa^* \Big)}  - \sqrt{2} {\Re\Big(T_{\lambda}\Big)} \Big)\nonumber\\ 
m_{\sigma_{u}\sigma_s} &= -\frac{1}{2} v_d \Big(2 v_s {\Re\Big(\lambda \kappa^* \Big)}  - \sqrt{2} {\Re\Big(T_{\lambda}\Big)} \Big)\nonumber\\ 
m_{\sigma_s\sigma_s} &= \frac{1}{2} \Big(-2 \sqrt{2} v_s {\Re\Big(T_{\kappa}\Big)}  + \Big(2 v_{s}^{2} \kappa  + v_d v_u \lambda \Big)\kappa^*  + \Big(\Big(v_{d}^{2} + v_{u}^{2}\Big)\lambda  + v_d v_u \kappa \Big)\lambda^* \Big) + m_S^2.
\label{eq:sigmamasses}
\end{align} 
with gauge-fixing contributions given by
\begin{align} 
m_{\xi,\sigma_{d}\sigma_{d}} &= \xi_Z \frac{1}{4} v_{d}^{2} \Big(g_1 \sin\Theta_W   + g_2 \cos\Theta_W  \Big)^{2} \nonumber\\ 
m_{\xi,\sigma_{d}\sigma_{u}} &= -\xi_Z  \frac{1}{4} v_d v_u \Big(g_1 \sin\Theta_W   + g_2 \cos\Theta_W  \Big)^{2}\nonumber \\ 
m_{\xi,\sigma_{u}\sigma_{u}} &= \xi_Z  \frac{1}{4} v_{u}^{2} \Big(g_1 \sin\Theta_W   + g_2 \cos\Theta_W  \Big)^{2} 
\end{align} 
The pseudo-scalar mass matrix is diagonalised by \(Z^A\): 
\begin{equation} 
Z^A m^2_{A^0} Z^{A,\dagger} = m^{dia}_{2,A^0}.
\end{equation} 
At the minimum of the tree-level potential one eigenvalue of this matrix, that of the would-be Goldstone boson, 
is given by $\xi M_Z$ which yields a massless state in Landau gauge.
However, this would cause problems
when calculating the two-loop corrections because the derivatives of some loop integrals diverge for vanishing masses \cite{Martin:2014bca,Elias-Miro:2014pca}. 
This problem is not present in the MSSM when working in the gaugeless limit (i.e. taking $g_1=g_2=0$)
because all derivatives of the Goldstone masses with respect of the VEVs
vanish. However, this is not generally the case, and is certainly not true for the NMSSM. Here we can make the preliminary rotation to isolate the would-be Goldstone boson $G^0$:
\begin{align}
\sigma_u =& \sigma_{ud} \cos\beta  -  G^0 \sin\beta,\qquad \sigma_d = \sigma_{ud} \sin \beta  + G^0\cos \beta  
\end{align}
with $\tan \beta = v_u/v_d\bigg|_{\rm{min}}$ (i.e. the ratio evaluated at the minimum of the potential). We can substitute this into equations (\ref{eq:sigmamasses}) and, when applying the tadpole equations (\ref{eq:tadpoleD} -- \ref{eq:tadpoleS}), find  $\xi M_Z$ for the mass of $G^0$. However, when we do not apply the tadpole equations, but take the derivatives of the Goldstone mass with respect to the VEVs we find non-zero values, for example the non-zero second derivatives \emph{in the gaugeless limit} are 
\begin{align}
 \frac{\partial^2 m_{G^0}^2}{\partial v_u^2}\bigg|_{\rm{min}} =& \lambda^2 \cos^2 \beta  \nonumber\\
\frac{\partial^2 m_{G^0}^2}{\partial v_d^2 \partial (G^0)^2}\bigg|_{\rm{min}} =& \lambda^2\sin^2 \beta  \nonumber\\
\frac{\partial^2 m_{G^0}^2}{\partial v_s^2 \partial (G^0)^2}\bigg|_{\rm{min}} =& \lambda^2 - \Re\Big(\lambda \kappa^* \Big)\sin 2 \beta.
\end{align}
These clearly never all vanish unless $\lambda = 0$.

However, in our approach, we solve the tadpole equations above for the minimum of the \emph{full} tree-level potential, and then calculate the masses of the scalars and pseudo-scalars \emph{in the gaugeless limit}. In this way, we find that the ``mass'' of the (now genuine) Goldstone boson, given by $\frac{\partial^2 V}{\partial (G^0)^2} $, is non-zero. We find (in the Landau gauge)
\begin{align}
\frac{\partial^2 V}{\partial (G^0)^2}\bigg|_{\rm Gaugeless,\ \rm{min}} =& - \frac{M_Z^2}{2} \cos^2 2 \beta,
\end{align}
which is always tachyonic, as we expect -- since it signifies that we are working away from the `true' minimum. In this way, by ensuring that the loop functions are evaluated correctly for tachyonic mass-squareds, we avoid the problem of massless scalars and thus avoid the problems of the ``Goldstone boson catastrophe'' \cite{Martin:2014bca,Elias-Miro:2014pca}, since in this case the Goldstone-boson mass entering into the loop functions does not vary with renormalisation scale (except for the small variation induced by corrections to $M_Z$). We stress that this solution is a valid and correct procedure when working in the gaugeless limit -- but only in that limit, so that if we took $g_1, g_2$ non-zero then we would have to solve the problem by other means.

\subsubsection{Loop-level scalar masses}
We have given above our conventions for the tree-level mass matrices for the neutral scalar and pseudo-scalar bosons; for the conventions of all other matrices we refer again to Ref.~\cite{Staub:2010ty}.
This reference contains also the full expressions for the one-loop self-energies $\Pi^{(1)}_{hh}(p^2) $
contributing to the CP-even mass matrix at the one-loop level. The focus of 
this work is on the corrections at the two-loop level $\Pi^{(2)}_{hh}(0)$. 
 At two loops the self-energies do not carry 
any momentum dependence because they are calculated in the effective potential approach. 
More details about the calculation 
are given in sec.~\ref{sec:method}.

Once the  one- and two-loop corrections have been calculated, the pole mass of the Higgs is then calculated by taking the real part of the poles
of the corresponding propagator matrices
\begin{equation}
\mathrm{Det}\left[ p^2_i \mathbf{1} - m^{2,h}(p^2) \right] = 0,
\label{eq:propagator}
\end{equation}
where
\begin{equation}
 m^{2,h}(p^2) = \tilde{m}^{2,h}_T -  \Pi^{(1)}_{hh}(p^2) - \Pi^{(2)}_{hh}(0)
\end{equation}
Here, \(\tilde{m}^2_{h,T}\) is the tree-level mass matrix from eq.\
(\ref{ScalarMass}). Equation (\ref{eq:propagator}) must be solved
for each eigenvalue $p^2=m^2_i$. 

\bibliographystyle{ArXiv}
\bibliography{lit}

\end{document}